\documentclass{article}
\usepackage{amssymb,amsmath,bm,hyperref}
\newtheorem{theorem}{Theorem}
\newtheorem{remark}{Remark}

\begin{document}

\title{Weyl's Lagrangian in teleparallel form}

\author{
 James Burnett\footnote{Electronic mail: J.Burnett@ucl.ac.uk} and
Dmitri Vassiliev\footnote{Electronic mail: D.Vassiliev@ucl.ac.uk}
\\
\textit{\small Department of Mathematics and Institute of Origins}
\\
\textit{\small University College London, Gower Street, London WC1E~6BT, UK}
}

\maketitle

\vspace{-25pt}

\begin{abstract}
The main result of the paper is a new representation for the Weyl
Lagrangian (massless Dirac Lagrangian). As the dynamical variable we use
the coframe, i.e. an orthonormal tetrad of covector fields.
We write down a simple Lagrangian -- wedge product of axial torsion with a
lightlike element of the coframe -- and show that this gives the Weyl Lagrangian
up to a nonlinear change of dynamical variable.
The advantage of our approach
is that it does not require the use of spinors, Pauli matrices or
covariant differentiation. The only geometric concepts we use are
those of a metric, differential form, wedge product and exterior derivative.
Our result assigns a variational meaning to the tetrad
representation of the Weyl equation suggested by J.~B.~Griffiths and R.~A.~Newing.
\end{abstract}

\newpage

\section{Main result}
\label{Main result}

Throughout this paper we work on a 4-manifold $M$ equipped with
prescribed Lorentzian metric $g$. The construction presented in the
paper is local so we do not make a priori assumptions on the
geometric structure of spacetime $\{M,g\}$. The metric $g$ is not
necessarily the Minkowski metric.

The accepted mathematical model
for a massless neutrino field is the following complex
linear partial differential equation on $M$ know as \emph{Weyl's equation}:
\begin{equation}
\label{Weyl's equation}
i\sigma^\alpha{}_{a\dot b}\{\nabla\}_\alpha\xi^a=0.
\end{equation}
The corresponding Lagrangian is
\begin{equation}
\label{Weyl's Lagrangian}
L_\mathrm{Weyl}(\xi):=
\frac i2
(\bar\xi^{\dot b}\sigma^\alpha{}_{a\dot b}\{\nabla\}_\alpha\xi^a
-
\xi^a\sigma^\alpha{}_{a\dot b}\{\nabla\}_\alpha\bar\xi^{\dot b})
*1.
\end{equation}
Here
$*1$ is the standard volume 4-form
(Hodge dual of the scalar 1),
$\sigma^\alpha$, $\alpha=0,1,2,3$, are Pauli matrices,
$\xi$ is the unknown 2-component spinor field
and $\{\nabla\}$ is the covariant derivative with
respect to the Levi-Civita connection
defined by formulae
(\ref{covariant derivative of a spinor field Levi-Civita}),
(\ref{Christoffel symbols}).

Throughout the paper we will often deal with a situation when a pair
of complex fields differs by a \emph{constant} complex factor of
modulus 1. We will say in this case that the two fields are equal modulo
$\mathrm{U}(1)$ and use the mathematical symbol
$\ \stackrel{\operatorname{mod}\mathrm{U}(1)}=\ $
to indicate this fact in formulae.

It is well known that Weyl's Lagrangian (\ref{Weyl's Lagrangian}) is
$\mathrm{U}(1)$-invariant:
\[
\xi\stackrel{\operatorname{mod}\mathrm{U}(1)}=\tilde\xi
\quad\implies\quad
L_\mathrm{Weyl}(\xi)=L_\mathrm{Weyl}(\tilde\xi).
\]
In view of this we call
two spinor fields equivalent if they are equal modulo
$\mathrm{U}(1)$
and gather spinor fields into equivalence classes
according to this relation.
We call an equivalence class of spinors \emph{nonvanishing} if its representatives
do not vanish at any point.

The purpose of our paper is to give an alternative, much simpler
and geometrically more transparent,
representation for the Weyl Lagrangian (\ref{Weyl's Lagrangian}).
To this end we introduce instead of the spinor field a different unknown
-- the so-called \emph{coframe}. A coframe is a quartet of real
covector fields
$\vartheta^j$, $j=0,1,2,3$,
satisfying the constraint
\begin{equation}
\label{constraint for coframe}
g=o_{jk}\,\vartheta^j\otimes\vartheta^k
\end{equation}
where
$o_{jk}=o^{jk}:=\operatorname{diag}(1,-1,-1,-1)$.
For the sake of clarity we repeat formula~(\ref{constraint for coframe})
giving the tensor indices explicitly and performing summation in
the frame indices explicitly:
$g_{\alpha\beta}=o_{jk}\,\vartheta^j_\alpha\vartheta^k_\beta
=\vartheta^0_\alpha\vartheta^0_\beta
-\vartheta^1_\alpha\vartheta^1_\beta
-\vartheta^2_\alpha\vartheta^2_\beta
-\vartheta^3_\alpha\vartheta^3_\beta$.

Formula (\ref{constraint for coframe}) means that
the coframe is a field of orthonormal bases with
orthonormality understood in the Lorentzian sense.
Of course, at every point of the manifold $M$ the choice of
coframe is not unique: there are 6 real degrees of freedom in
choosing the coframe and any pair of coframes is related by a
Lorentz transformation.

At a physical level choosing the coframe as the unknown quantity
means that we allow every point of spacetime to rotate and assume
that rotations of different points are totally independent. These
rotations are described mathematically by attaching to each
spacetime point a coframe (=~orthonormal basis). The approach in which
the coframe plays the role of the dynamical variable is known as
\emph{teleparallelism} (=~absolute parallelism). This is a subject
promoted by A.~Einstein and \'E.~Cartan \cite{letters,unzicker-2005-,sauer}.

The idea of rotating points may seem exotic, however it has long
been accepted in continuum mechanics within the so-called Cosserat
theory of elasticity \cite{Co}. The Cosserat theory of elasticity
has been in existence since 1909 and appears under various names in
modern applied mathematics literature such as \emph{oriented
medium}, \emph{asymmetric elasticity}, \emph{micropolar elasticity},
\emph{micromorphic elasticity}, \emph{moment elasticity} etc.
Cosserat elasticity is closely related to the theory of
ferromagnetic materials \cite{Ba} and the theory of liquid crystals
\cite{Lin,Ba2}. As to teleparallelism, it is, effectively, a special
case of Cosserat elasticity: here the assumption is that the elastic
continuum experiences no displacements, only rotations. With regards
to the latter it is interesting that Cartan acknowledged
\cite{cartan1} that he drew inspiration from the monograph \cite{Co}
of the Cosserat brothers.

Define the 3-form
\begin{equation}
\label{axial torsion}
T^\mathrm{ax}:=\frac13\,o_{jk}\,\vartheta^j\wedge d\vartheta^k
\end{equation}
where $\,d\,$ denotes the exterior derivative.
This 3-form is called \emph{axial torsion of the teleparallel
connection}. The geometric meaning of the latter phrase is
explained in a concise fashion in
Appendix \ref{A brief introduction to teleparallelism}
whereas
a detailed exposition of the application of torsion in field theory
and the history of the subject
can be found in \cite{goenner,cartantorsionreview}.
What is important at this stage is the
observation that the 3-form (\ref{axial torsion}) is a measure of
deformations generated by rotations of spacetime points.

Note that the 3-form (\ref{axial torsion}) has the remarkable
property of conformal covariance: if we rescale our metric and
coframe as
\begin{equation}
\label{rescaling metric}
g_{\alpha\beta}\mapsto e^{2h}g_{\alpha\beta}
\end{equation}
\begin{equation}
\label{rescaling coframe}
\vartheta^j\mapsto e^h\vartheta^j
\end{equation}
where $h:M\to\mathbb{R}$ is an arbitrary scalar function,
then our 3-form is scaled as
\begin{equation}
\label{rescaling axial torsion}
T^\mathrm{ax}\mapsto e^{2h}T^\mathrm{ax}
\end{equation}
without the derivatives of $h$ appearing. The issue of conformal
covariance and invariance will be examined in detail in Section
\ref{Conformal invariance}.

It is tempting to use the 3-form (\ref{axial torsion}) as our
Lagrangian but the problem is that we are working in 4-space. In
order to turn our 3-form into a 4-form we proceed as follows.

Put
\begin{equation}
\label{definition of covector field l}
l:=\vartheta^0+\vartheta^3.
\end{equation}
This is a a nonvanishing real lightlike covector field. It will
eventually (see Section \ref{Discussion})
transpire that the covector field
(\ref{definition of covector field l}) has the geometric meaning of
neutrino current.

We define our ``teleparallel'' Lagrangian as
\begin{equation}
\label{teleparallel Lagrangian}
L_\mathrm{tele}(\vartheta):=l\wedge T^\mathrm{ax}.
\end{equation}
Note that formulae
(\ref{axial torsion}),
(\ref{definition of covector field l}),
(\ref{teleparallel Lagrangian})
are very simple.
They do not contain spinors, Pauli matrices or covariant derivatives.
The only concepts used are those of a differential form, wedge product
and exterior derivative. Even the metric does not appear in formulae
(\ref{axial torsion}),
(\ref{definition of covector field l}),
(\ref{teleparallel Lagrangian})
explicitly: it is incorporated implicitly via
the constraint (\ref{constraint for coframe}).

Let us now examine the behaviour of our Lagrangian
(\ref{teleparallel Lagrangian}) under Lorentz transformations of the
coframe:
\begin{equation}
\label{Lorentz transformation 1}
\vartheta^j
\stackrel\Lambda\mapsto\tilde\vartheta^j=\Lambda^j{}_k\vartheta^k
\end{equation}
where the $\Lambda^j{}_k$ are real scalar functions satisfying the constraint
\begin{equation}
\label{Lorentz transformation 2}
o_{jk}\Lambda^j{}_r\Lambda^k{}_s=o_{rs}.
\end{equation}
Obviously, transformations (\ref{Lorentz transformation 1}),
(\ref{Lorentz transformation 2}) form an infinite-dimensional Lie
group.
Within this group we single out an
infinite-dimensional Lie subgroup $H$ as follows.
Put
\begin{equation}
\label{definition of covector field m}
m:=\vartheta^1+i\vartheta^2.
\end{equation}
The subgroup $H$ is defined by the condition of preservation
modulo $\mathrm{U}(1)$
of the complex 2-form $l\wedge m$. More precisely,
a Lorentz transformation (\ref{Lorentz transformation 1}),
(\ref{Lorentz transformation 2}) is included in $H$ if and only if
\begin{equation}
\label{definition of subgroup H}
l\wedge m
\stackrel{\operatorname{mod}\mathrm{U}(1)}=
\tilde l\wedge\tilde m
\end{equation}
where $\tilde l=\tilde\vartheta^0+\tilde\vartheta^3$
and $\tilde m=\tilde\vartheta^1+i\tilde\vartheta^2$.

The first main result of our paper is

\begin{theorem}
\label{main theorem 1} The teleparallel Lagrangian
(\ref{teleparallel Lagrangian}) is invariant under the action of the
group $H$.
\end{theorem}

In view of Theorem \ref{main theorem 1} we call two coframes
equivalent if they differ by a transformation from the subgroup $H$
and gather coframes into equivalence classes according to this
relation.

The second main result of our paper is

\begin{theorem}
\label{main theorem 2}
The equivalence classes of coframes $\vartheta$ and nonvanishing spinor fields
$\xi$ are in a one-to-one correspondence given by the formula
\begin{equation}
\label{relationship between coframe and spinor field}
\left(l\wedge m\right)_{\alpha\beta}
\stackrel{\operatorname{mod}\mathrm{U}(1)}=
\sigma_{\alpha\beta ab}\xi^a\xi^b
\end{equation}
where $l$ and $m$ are defined by formulae
(\ref{definition of covector field l})
and
(\ref{definition of covector field m})
respectively,
$\vartheta$ and $\xi$ are arbitrary representatives of
the corresponding equivalence classes and
$\sigma_{\alpha\beta}$ are ``second order'' Pauli matrices
(\ref{second order Pauli matrices}).
Furthermore, under the
correspondence~(\ref{relationship between coframe and spinor field}) we have
\begin{equation}
\label{relation between two Lagrangians}
L_\mathrm{tele}(\vartheta)=-\frac43\,L_\mathrm{Weyl}(\xi).
\end{equation}
\end{theorem}

A shorter way of stating Theorem
\ref{main theorem 2} is ``the nonlinear change of variable
\[
\text{coframe}\ \vartheta
\quad\longleftrightarrow\quad
\text{spinor field}\ \xi
\]
specified by formula
(\ref{relationship between coframe and spinor field})
shows that the two Lagrangians, $L_\mathrm{tele}(\vartheta)$
and $L_\mathrm{Weyl}(\xi)$, are the same up to a constant factor''.
The only problem with such a statement is that it
brushes under the carpet the important question of gauge invariance.

The above results were announced, without a detailed proof, in
the short communication \cite{vassilievPRD}.

The paper has the following structure.
In Section~\ref{The gauge group H}
we describe explicitly the gauge group $H$
which we initially defined implicitly
by formula (\ref{definition of subgroup H}).
In Sections~\ref{Proof of Theorem 1} and \ref{Proof of Theorem 2}
we prove Theorems \ref{main theorem 1} and \ref{main theorem 2} respectively.
In Section~\ref{Conformal invariance}
we present a modified version of our construction
which makes conformal invariance more obvious.
The concluding discussion is contained in Section~\ref{Discussion}.

\section{Notation and conventions}
\label{Notation and conventions}

Our notation follows \cite{vassilievPRD,MR1841284,MR1925542,MR2132536,MR2176749}.
In particular, in line with the traditions of particle physics,
we use Greek letters to denote tensor (holonomic) indices.
We identify differential forms with antisymmetric tensors.

All our constructions are local and occur in a neighbourhood of a
given point $P\in M$. Moreover, we assume that we have a
given reference coframe ${\bm{\vartheta}}$ defined in a
neighbourhood of $P$; we need this reference coframe to specify
orientation and positive direction of time.

We restrict our choice of local coordinates on $M$ to those with
$\det{\bm{\vartheta}}{}^j_\alpha>0$. This means that we work in
local coordinates with specific orientation. In particular, this
allows us to define the Hodge star:
we define the action of $*$ on a rank $r$
antisymmetric tensor $R$ as
\begin{equation}
\label{definition of Hodge star}
(*R)_{\alpha_{r+1}\ldots\alpha_4}:=(r!)^{-1}\,\sqrt{|\det g|}\,
R^{\alpha_1\ldots\alpha_r}\varepsilon_{\alpha_1\ldots\alpha_4}
\end{equation}
where $\varepsilon$ is the totally antisymmetric quantity,
$\varepsilon_{0123}:=+1$.

The coframe $\vartheta$ which serves as our dynamical variable is
assumed to satisfy
\begin{equation}
\label{orientation of coframe}
\det\vartheta^j_\alpha>0,
\end{equation}
and ${\bm{\vartheta}}{}^0\cdot\vartheta^0>0$.
These assumptions mean that we work with coframes
$\vartheta$ which can be obtained from our reference coframe
${\bm{\vartheta}}$ by proper Lorentz transformations:
$
\vartheta^j=\Lambda^j{}_k\,{\bm{\vartheta}}{}^k
$
where the $\Lambda^j{}_k$ are real scalar functions satisfying conditions
\[
o_{ji}\,\Lambda^j{}_k\,\Lambda^i{}_r=o_{kr},
\qquad
\det\Lambda^j{}_k>0,
\qquad
\Lambda^0{}_0>0.
\]

We define the forward light cone (at a given point) as the set of covectors
of the form $c_j\vartheta^j$ with $o^{jk}c_jc_k=0$ and $c_0>0$.
This implies, in particular, that our covector $l$ defined by formula
(\ref{definition of covector field l})
lies on the forward light cone.

Details of our spinor notation are given in
Appendix A of \cite{MR2176749}.
In particular, the defining relation for Pauli matrices is
$\sigma^\alpha{}_{a\dot b}\sigma^{\beta c\dot b}
+\sigma^\beta{}_{a\dot b}\sigma^{\alpha c\dot b}
=2g^{\alpha\beta}\delta_a{}^c$.

Consider (at a given point) covectors of the form
$\sigma_{\alpha a\dot b}\xi^a\bar\xi^{\dot b}$, $\xi\ne0$.
These covectors are
lightlike and the set of all such covectors forms a cone. We assume
that this cone is the forward light cone defined above. In other
words, we assume that the positive direction of time encoded in our
Pauli matrices agrees with the positive direction of time encoded in
our coframe.

We define
\begin{equation}
\label{second order Pauli matrices}
\sigma_{\alpha\beta ac}:=(1/2)
(
\sigma_{\alpha a\dot b}\epsilon^{\dot b\dot d}\sigma_{\beta c\dot d}
-
\sigma_{\beta a\dot b}\epsilon^{\dot b\dot d}\sigma_{\alpha c\dot d}
)
\end{equation}
where
\[
\epsilon_{ab}=\epsilon_{\dot a\dot b}=
\epsilon^{ab}=\epsilon^{\dot a\dot b}=
\begin{pmatrix}
0&1\\
-1&0
\end{pmatrix}
\]
(the first spinor index enumerates the rows and the second one the
columns). These ``second order'' Pauli matrices are polarized, i.e.
$*\sigma=\pm i\sigma$ depending on the choice of ``basic'' Pauli
matrices $\sigma_{\alpha a\dot b}\,$. Here the explicit formula for
the action of the Hodge star on second order Pauli matrices is
\[
(*\sigma)_{\gamma\delta ab}:=\frac12\,\sqrt{|\det g|}
\ \sigma^{\alpha\beta}{}_{ab}\,\varepsilon_{\alpha\beta\gamma\delta}.
\]
We assume that
\begin{equation}
\label{polarization of Pauli matrices}
*\sigma=-i\sigma.
\end{equation}

Note that formula (\ref{orientation of coframe}) implies
\begin{equation}
\label{polarization of l wedge m}
*(l\wedge m)=-i(l\wedge m)
\end{equation}
where the covectors $l$ and $m$ are defined by formulae
(\ref{definition of covector field l})
and
(\ref{definition of covector field m})
respectively.
We chose the sign in the RHS of
(\ref{polarization of Pauli matrices}) so as to agree with
(\ref{polarization of l wedge m}).
In other words, the meaning of condition (\ref{polarization of Pauli matrices})
is that the orientation encoded in our Pauli
matrices agrees with the orientation encoded in our coframe.

The covariant derivatives of a vector field and a spinor field are
defined as
\begin{equation}
\label{covariant derivative of a vector field}
\nabla_\alpha v^\beta:=\partial_\alpha v^\beta
+\Gamma^\beta{}_{\alpha\gamma}v^\gamma,
\end{equation}
\begin{equation}
\label{covariant derivative of a spinor field}
\nabla_\alpha\xi^a:=
\partial_\alpha\xi^a
+\frac14\sigma_\beta{}^{a\dot c}
(\partial_\alpha\sigma^\beta{}_{b\dot c}
+\Gamma^\beta{}_{\alpha\gamma}\sigma^\gamma{}_{b\dot c})\xi^b
\end{equation}
where $\Gamma^\beta{}_{\alpha\gamma}$ are the connection coefficients.
Throughout the main text of the paper we use the Levi-Civita connection and
indicate this by curly brackets. That is, for the
Levi-Civita connection we write formulae
(\ref{covariant derivative of a vector field}),
(\ref{covariant derivative of a spinor field}) as
\begin{equation}
\label{covariant derivative of a vector field Levi-Civita}
\{\nabla\}_\alpha v^\beta:=\partial_\alpha v^\beta
+\{\Gamma\}^\beta{}_{\alpha\gamma}v^\gamma,
\end{equation}
\begin{equation}
\label{covariant derivative of a spinor field Levi-Civita}
\{\nabla\}_\alpha\xi^a:=
\partial_\alpha\xi^a
+\frac14\sigma_\beta{}^{a\dot c}
(\partial_\alpha\sigma^\beta{}_{b\dot c}
+\{\Gamma\}^\beta{}_{\alpha\gamma}\sigma^\gamma{}_{b\dot c})\xi^b
\end{equation}
where
\begin{equation}
\label{Christoffel symbols}
\{\Gamma\}^\beta{}_{\alpha\gamma}=
\left\{{{\beta}\atop{\alpha\gamma}}\right\}:=
\frac12g^{\beta\delta}
(\partial_\alpha g_{\gamma\delta}
+\partial_\gamma g_{\alpha\delta}
-\partial_\delta g_{\alpha\gamma})
\end{equation}
are the Christoffel symbols uniquely determined by the metric.
An alternative (teleparallel) connection will be introduced in
Appendix \ref{A brief introduction to teleparallelism}.

In performing subsequent calculations
it will be convenient for us to switch from the real coframe
$(\vartheta^0,\vartheta^1,\vartheta^2,\vartheta^3)$ to the complex
coframe $(l,m,n)$ where $l$, $m$ and $n$ are given by formulae
(\ref{definition of covector field l}),
(\ref{definition of covector field m})
and
\begin{equation}
\label{definition of covector field n}
n:=\vartheta^0-\vartheta^3
\end{equation}
respectively.
Note that in this new notation
the constraint (\ref{constraint for coframe}) takes the form
\begin{equation}
\label{constraint for coframe in complex form}
g=(1/2)(l\otimes n+n\otimes l-m\otimes\bar m-\bar m\otimes m).
\end{equation}
The quartet of covectors $(l,m,\bar m,n)$ is known as a
\emph{null tetrad} or a
\emph{Newman--Penrose tetrad}~\cite{newmanpenrose}.

\section{The gauge group $H$}
\label{The gauge group H}

In this section
we describe explicitly the gauge group $H$
which we initially defined implicitly
by formula (\ref{definition of subgroup H}).

Consider a Lorentz transformation of the coframe
(\ref{Lorentz transformation 1}) satisfying the defining condition
(\ref{definition of subgroup H}) of our group $H$. (Recall that here
the $\Lambda^j{}_k$ are not assumed to be constant, i.e. they are
real scalar functions satisfying (\ref{Lorentz transformation 2}).)
We denote this Lorentz transformation $\Lambda$.

Condition (\ref{definition of subgroup H}) means that $\Lambda$ is a
composition of two Lorentz transformations:
\begin{equation}
\label{The gauge group H equation 1}
\Lambda=\Lambda''\Lambda'
\end{equation}
where $\Lambda'$ is a rotation by a constant angle
$\varphi$ in the $\vartheta^1,\vartheta^2$--plane
\begin{equation}
\label{The gauge group H equation 2}
\begin{pmatrix}
l\\ m\\ n
\end{pmatrix}
\stackrel{\Lambda'}\mapsto
\begin{pmatrix}
l\\ e^{i\varphi}m\\ n
\end{pmatrix}
\end{equation}
and $\Lambda''$ is a Lorentz transformation preserving the 2-form
$l\wedge m$.
Our convention for writing compositions of Lorentz transformations is as follows.
When looking at a Lorentz transformation (\ref{Lorentz transformation 1})
we view the real coframe as a column of height~4
with entries $\vartheta^k$, $k=0,1,2,3$,
and the the Lorentz transformation itself
as multiplication by a real $4\times 4$~matrix $\Lambda^j{}_k$,
so the group
operation is matrix multiplication with the matrix furthest to the right
acting on the coframe first.
Say, formula (\ref{The gauge group H equation 1}) means that $\Lambda'$
acts on the coframe first.

It is known, see Section 10.122 in \cite{Besse}, that
Lorentz transformations preserving the 2-form $l\wedge m$
admit an explicit description:
\begin{equation}
\label{The gauge group H equation 3}
\begin{pmatrix}
l\\ m\\ n
\end{pmatrix}
\stackrel{\Lambda''}\mapsto
\begin{pmatrix}
l\\ m+fl\\ n+f\bar m+\bar fm+|f|^2l
\end{pmatrix}
\end{equation}
where $f:M\to\mathbb{C}$ is an arbitrary scalar function.
Substituting
(\ref{The gauge group H equation 2}),
(\ref{The gauge group H equation 3})
into (\ref{The gauge group H equation 1})
we arrive at the explicit formula
for an element $\Lambda$ of the group $H$:
\begin{equation}
\label{The gauge group H equation 4}
\begin{pmatrix}
l\\ m\\ n
\end{pmatrix}
\stackrel\Lambda\mapsto
\begin{pmatrix}
l\\ e^{i\varphi}m+fl\\ n+fe^{-i\varphi}\bar m+\bar fe^{i\varphi}m+|f|^2l
\end{pmatrix}.
\end{equation}

Let us now examine the structure of the group $H$.

The group of rotations
in the $\vartheta^1,\vartheta^2$--plane
is isomorphic to $\mathrm{U}(1)$.
Hence further on we will refer to the group of Lorentz transformations
of the coframe of the form
(\ref{The gauge group H equation 2})
as $\mathrm{U}(1)$.
Let us emphasise that the $\varphi$ appearing in formula
(\ref{The gauge group H equation 2})
is a constant, not a function.

Let us denote by $B^2(M)$ the group of Lorentz transformations
of the coframe preserving the 2-form $l\wedge m$, see formula
(\ref{The gauge group H equation 3}).
In choosing the notation $B^2$ we follow
\cite{Besse} whereas the
``$M$'' indicates dependence on the point of the manifold $M$,
i.e. it highlights the fact that the $f$
appearing in formula (\ref{The gauge group H equation 3})
is a function, not a constant.

Both $\mathrm{U}(1)$ and $B^2(M)$ are
abelian\footnote
{The group $B^2$ can, in fact, be characterised as the nontrivial abelian subgroup
of the Lorentz group. See Appendix B in \cite{vassilievPRD} for details.}
subgroups of $H$.
Moreover, it is easy to see that $B^2(M)$ is a normal subgroup of $H$,
$B^2(M)\triangleleft H$, and that $H$ is a semidirect product of
$B^2(M)$ and $\mathrm{U}(1)$, $H=B^2(M)\ltimes\mathrm{U}(1)$.

The infinite-dimensional Lie group $H$ is itself nonabelian.
However, it is
very close to being abelian: $H$ contains the
infinite-dimensional abelian Lie subgroup
$B^2(M)$ of codimension 1.

\section{Proof of Theorem \ref{main theorem 1}}
\label{Proof of Theorem 1}

Let us rewrite our teleparallel Lagrangian (\ref{teleparallel Lagrangian})
in terms of the complex coframe
(\ref{definition of covector field l}),
(\ref{definition of covector field m}),
(\ref{definition of covector field n}):
\begin{equation}
\label{Lagrangian in complex form}
L_\mathrm{tele}(\vartheta)
=(1/6)\ l\wedge(n\wedge dl-\bar m\wedge dm-m\wedge d\bar m).
\end{equation}
The group $H$ is a semidirect product of the groups
$B^2(M)$ and $\mathrm{U}(1)$ so in order to check that
(\ref{Lagrangian in complex form}) is invariant under the action of
$H$ it is sufficient to check that (\ref{Lagrangian in complex form})
is invariant under the actions of $B^2(M)$ and $\mathrm{U}(1)$ separately.
\linebreak
$\mathrm{U}(1)$-invariance is obvious: just substitute
(\ref{The gauge group H equation 2}) into (\ref{Lagrangian in complex form})
noting that $\varphi$ is constant. Hence, it remains only to
we check that our teleparallel Lagrangian (\ref{Lagrangian in complex form})
is invariant under the transformation (\ref{The gauge group H equation 3}).

When substituting (\ref{The gauge group H equation 3}) into
(\ref{Lagrangian in complex form}) we will get
an expression which is a sum of two terms:
\begin{itemize}
\item
term without derivatives of the function $f$, and
\item
term with derivatives of the function $f$.
\end{itemize}
Looking at our original formula (\ref{teleparallel Lagrangian}) we see that
the term without derivatives of the function $f$ does not change the
teleparallel Lagrangian because
our transformation (\ref{The gauge group H equation 3}) preserves the covector
field $l$ and because axial torsion is an irreducible piece of torsion
(i.e. the 3-form (\ref{axial torsion}) is invariant under rigid
Lorentz transformations). So it only remains to check that the term with
derivatives of the function $f$ vanishes. The term in question is
\[
(1/6)\ l\wedge(-\bar m\wedge df\wedge l-m\wedge d\bar f\wedge l)
\]
which is clearly zero. $\square$

\section{Proof of Theorem \ref{main theorem 2}}
\label{Proof of Theorem 2}

The gauge group $H$ allows us to gather coframes into equivalence classes:
we call two coframes equivalent if they differ by a transformation
from $H$. We will now establish the geometric meaning of these
equivalence classes of coframes.

Let us first fix a spacetime point $x\in M$
and examine in detail the geometric meaning of the group $B^2$.
We initially defined $B^2$ as the the group of Lorentz transformations
preserving the 2-form $l\wedge m$. The complex nonzero antisymmetric tensor
$l\wedge m$ is polarized (see (\ref{polarization of l wedge m}))
and has the additional property $\det(l\wedge m)=0$. It is easy to see
(and this fact was extensively used in
\cite{vassilievPRD,MR1841284,MR1925542,MR2132536,MR2176749})
that such a tensor can be written
in terms of a nonzero spinor $\xi$ as
\begin{equation}
\label{Proof of Theorem 2 equation 1}
\left(l\wedge m\right)_{\alpha\beta}
=
\sigma_{\alpha\beta ab}\xi^a\xi^b
\end{equation}
with the spinor defined uniquely up to sign.
Thus, the group $B^2$ can be reinterpreted as
the group of Lorentz transformations
preserving a given nonzero spinor $\xi$ and the equivalence
classes of coframes are related to this spinor according to formula
(\ref{Proof of Theorem 2 equation 1}).
Here the relationship between an equivalence class of coframes and a nonzero spinor
is one-to-two because formula (\ref{Proof of Theorem 2 equation 1}) allows
us to change the sign of $\xi$.

\begin{remark}
One can use the above observation to formulate an alternative
definition of a spinor: a spinor is a coset of the Lorentz group
with respect to the subgroup $B^2$. In using this definition one,
however, has to decide whether to use left or right cosets as $B^2$
is not a normal subgroup of the Lorentz group.
\end{remark}

\begin{remark}
In $\mathrm{SL}(2,\mathbb{C})$ notation the group $B^2$ is written
in a particularly simple way:
$B^2=\left\{
\left.
\left(
\begin{matrix}
\ 1\ &\ f\ \\
\ 0\ &\ 1\
\end{matrix}
\right)
\right|
\quad f\in\mathbb{C}
\right\}$.
\end{remark}

Let us now allow dependence on the spacetime point $x\in M$.
Then the group $B^2(M)$ is the group of Lorentz transformations
preserving a given nonzero spinor field $\xi$, with the equivalence
classes of coframes related to the spinor field according to formula
(\ref{Proof of Theorem 2 equation 1}).
Here the relationship between an equivalence class of coframes and
a nonvanishing spinor field remains one-to-two.

Finally, let us switch from the group $B^2(M)$ to
$H=B^2(M)\ltimes\mathrm{U}(1)$. This means that in our definition of
equivalence classes of coframes we allow $l\wedge m$ to be
multiplied by a constant complex factor of modulus 1, so formula
(\ref{Proof of Theorem 2 equation 1}) turns into
(\ref{relationship between coframe and spinor field}).
Here the relationship between an
equivalence class of coframes and a nonvanishing spinor field
becomes one-to-infinity because formula
(\ref{relationship between coframe and spinor field})
allows us to multiply the nonvanishing
spinor field $\xi$ by a constant complex factor of modulus 1;
note that this eliminates the difference between $\xi$ and $-\xi$. It
remains only to gather nonvanishing spinor fields $\xi$ into
equivalence classes as described in the beginning of Section \ref{Main result}
and we arrive at a one-to-one correspondence
between equivalence classes of coframes and nonvanishing spinor fields
given by the explicit formula
(\ref{relationship between coframe and spinor field}).

In the remainder of this section we perform the nonlinear
change of variable
\[
\text{spinor field}\ \xi
\quad\longrightarrow\quad
\text{coframe}\ \vartheta
\]
and show that $L_\mathrm{Weyl}(\xi)$ turns into
$-\frac34\,L_\mathrm{tele}(\vartheta)$.
In order to simplify calculations we observe
that we have freedom in our choice of Pauli matrices. It is sufficient
to prove formula (\ref{relation between two Lagrangians}) for one particular
choice of Pauli matrices, hence it is natural to choose Pauli matrices in
a way that makes calculations as simple as possible.
Note that this trick is not new: it was, for example, extensively used by
A.~Dimakis and F.~M\"uller-Hoissen
\cite{muellerhoissen1,muellerhoissen2,muellerhoissen3}.

We choose Pauli matrices
\begin{equation}
\label{special formula for Pauli matrices}
\sigma_{\alpha a\dot b}=\vartheta^j_\alpha\,s_{ja\dot b}
\end{equation}
where
\begin{equation}
\label{Pauli matrices s}
s_{ja\dot b}=
\begin{pmatrix}
s_{0a\dot b}\\
s_{1a\dot b}\\
s_{2a\dot b}\\
s_{3a\dot b}
\end{pmatrix}
:=
\begin{pmatrix}
\begin{pmatrix}
1&0\\
0&1
\end{pmatrix}
\\
\begin{pmatrix}
0&1\\
1&0
\end{pmatrix}
\\
\begin{pmatrix}
0&i\\
-i&0
\end{pmatrix}
\\
\begin{pmatrix}
1&0\\
0&-1\end{pmatrix}
\end{pmatrix}.
\end{equation}
Let us stress that in the statement of Theorem \ref{main theorem 2}
Pauli matrices are not assumed to be related in any way to the
coframe $\vartheta$. We are just choosing the particular Pauli
matrices
(\ref{special formula for Pauli matrices}),
(\ref{Pauli matrices s})
to simplify calculations in our proof.

Note that the matrices
(\ref{special formula for Pauli matrices}),
(\ref{Pauli matrices s})
satisfy all the conditions listed in Section~\ref{Notation and conventions}.

We now calculate explicitly the corresponding second order Pauli matrices:
\begin{equation}
\label{special formula for second order Pauli matrices}
\sigma_{\alpha\beta ab}=\frac12(\vartheta^j\wedge\vartheta^k)_{\alpha\beta}
\,s_{jkab}
\end{equation}
where
\begin{multline}
\label{second order Pauli matrices s}
s_{jkab}=
\begin{pmatrix}
0&s_{01ab}&s_{02ab}&s_{03ab}\\
s_{10ab}&0&s_{12ab}&s_{13ab}\\
s_{20ab}&s_{21ab}&0&s_{23ab}\\
s_{30ab}&s_{31ab}&s_{32ab}&0
\end{pmatrix}
\\
:=
\begin{pmatrix}
0&
\begin{pmatrix}
1&0\\
0&-1
\end{pmatrix}&
\begin{pmatrix}
i&0\\
0&i
\end{pmatrix}&
\begin{pmatrix}
0&-1\\
-1&0
\end{pmatrix}\\
\begin{pmatrix}
-1&0\\
0&1
\end{pmatrix}&
0&
\begin{pmatrix}
0&i\\
i&0
\end{pmatrix}&
\begin{pmatrix}
-1&0\\
0&-1
\end{pmatrix}\\
\begin{pmatrix}
-i&0\\
0&-i
\end{pmatrix}&
\begin{pmatrix}
0&-i\\
-i&0
\end{pmatrix}&
0&
\begin{pmatrix}
-i&0\\
0&i
\end{pmatrix}\\
\begin{pmatrix}
0&1\\
1&0
\end{pmatrix}&
\begin{pmatrix}
1&0\\
0&1
\end{pmatrix}&
\begin{pmatrix}
i&0\\
0&-i
\end{pmatrix}&
0
\end{pmatrix}.
\end{multline}
Substituting
(\ref{definition of covector field l}),
(\ref{definition of covector field m})
and
(\ref{special formula for second order Pauli matrices}),
(\ref{second order Pauli matrices s})
into the equation
(\ref{relationship between coframe and spinor field})
we see that this equation can be easily resolved
for $\xi$ giving
\begin{equation}
\label{special formula for spinor}
\xi^a
\stackrel{\operatorname{mod}\mathrm{U}(1)}=
\begin{pmatrix}
1\\
0
\end{pmatrix}.
\end{equation}
Formula (\ref{special formula for spinor}) may seem strange: we are proving
Theorem \ref{main theorem 2} for a general nonvanishing spinor field $\xi$
but ended up with formula (\ref{special formula for spinor}) which is very specific.
However, there is no contradiction here because we chose Pauli matrices
specially adapted to the coframe and, hence, specially adapted
to the corresponding spinor field.

Substituting (\ref{covariant derivative of a spinor field Levi-Civita})
and
(\ref{special formula for spinor}) into
(\ref{Weyl's Lagrangian}) we get
\begin{multline*}
L_\mathrm{Weyl}(\xi)
\\
=\frac i8
(\bar\xi^{\dot b}\sigma^\alpha{}_{a\dot b}
\sigma_\beta{}^{a\dot c}
(\partial_\alpha\sigma^\beta{}_{d\dot c}
+\{\Gamma\}^\beta{}_{\alpha\gamma}\sigma^\gamma{}_{d\dot c})\xi^d
-
\xi^a\sigma^\alpha{}_{a\dot b}
\sigma_\beta{}^{c\dot b}
(\partial_\alpha\sigma^\beta{}_{c\dot d}
+\{\Gamma\}^\beta{}_{\alpha\gamma}\sigma^\gamma{}_{c\dot d})\bar\xi^{\dot d})
*1
\\
=\frac i8
(\sigma^\alpha{}_{a\dot 1}
\sigma_\beta{}^{a\dot c}
(\partial_\alpha\sigma^\beta{}_{1\dot c}
+\{\Gamma\}^\beta{}_{\alpha\gamma}\sigma^\gamma{}_{1\dot c})
-
\sigma^\alpha{}_{1\dot b}
\sigma_\beta{}^{c\dot b}
(\partial_\alpha\sigma^\beta{}_{c\dot 1}
+\{\Gamma\}^\beta{}_{\alpha\gamma}\sigma^\gamma{}_{c\dot 1}))
*1
\\
=\frac i8
(\sigma^\alpha{}_{a\dot 1}
\sigma_\beta{}^{a\dot c}
\{\nabla\}_\alpha\sigma^\beta{}_{1\dot c}
-
\sigma^\alpha{}_{1\dot b}
\sigma_\beta{}^{c\dot b}
\{\nabla\}_\alpha\sigma^\beta{}_{c\dot 1})
*1\,.
\end{multline*}
We now write down the spinor summation indices explicitly:
\begin{multline*}
L_\mathrm{Weyl}(\xi)
=\frac i8(
\sigma^\alpha{}_{1\dot 1}
\sigma_\beta{}^{1\dot 2}
\{\nabla\}_\alpha\sigma^\beta{}_{1\dot 2}
+\sigma^\alpha{}_{2\dot 1}
\sigma_\beta{}^{2\dot 1}
\{\nabla\}_\alpha\sigma^\beta{}_{1\dot 1}
+\sigma^\alpha{}_{2\dot 1}
\sigma_\beta{}^{2\dot 2}
\{\nabla\}_\alpha\sigma^\beta{}_{1\dot 2}
\\
-\sigma^\alpha{}_{1\dot 1}
\sigma_\beta{}^{2\dot 1}
\{\nabla\}_\alpha\sigma^\beta{}_{2\dot 1}
-\sigma^\alpha{}_{1\dot 2}
\sigma_\beta{}^{1\dot 2}
\{\nabla\}_\alpha\sigma^\beta{}_{1\dot 1}
-\sigma^\alpha{}_{1\dot 2}
\sigma_\beta{}^{2\dot 2}
\{\nabla\}_\alpha\sigma^\beta{}_{2\dot 1}
)*1\,.
\end{multline*}
Note that the terms with $a=1$, $\dot c=\dot 1$ and
$\dot b=\dot 1$, $c=1$ cancelled out.
Finally, we substitute explicit formulae
(\ref{special formula for Pauli matrices}),
(\ref{Pauli matrices s})
for our Pauli matrices which gives us
\begin{multline*}
L_\mathrm{Weyl}(\xi)
=\frac i8(
l^\alpha
(-\bar m_\beta)
\{\nabla\}_\alpha m^\beta
+\bar m^\alpha
(-m_\beta)
\{\nabla\}_\alpha l^\beta
+\bar m^\alpha
l_\beta
\{\nabla\}_\alpha m^\beta
\\
-l^\alpha
(-m_\beta)
\{\nabla\}_\alpha\bar m ^\beta
-m^\alpha
(-\bar m_\beta)
\{\nabla\}_\alpha l^\beta
-m^\alpha
l_\beta
\{\nabla\}_\alpha\bar m^\beta
)*1
\\
=\frac i8(
(m\wedge\bar m)^{\alpha\beta}\{\nabla\}_\alpha l_\beta
-(l\wedge\bar m)^{\alpha\beta}\{\nabla\}_\alpha m_\beta
+(l\wedge m)^{\alpha\beta}\{\nabla\}_\alpha\bar m_\beta
)*1
\\
=\frac i{16}(
(m\wedge\bar m)^{\alpha\beta}(dl)_{\alpha\beta}
-(l\wedge\bar m)^{\alpha\beta}(dm)_{\alpha\beta}
+(l\wedge m)^{\alpha\beta}(d\bar m)_{\alpha\beta}
)*1
\\
=\frac i{16}*(
(m\wedge\bar m)^{\alpha\beta}(dl)_{\alpha\beta}
-(l\wedge\bar m)^{\alpha\beta}(dm)_{\alpha\beta}
+(l\wedge m)^{\alpha\beta}(d\bar m)_{\alpha\beta}
)
\\
=\frac i8(
[*(m\wedge\bar m)]\wedge dl
-[*(l\wedge\bar m)]\wedge dm
+[*(l\wedge m)]\wedge d\bar m
).
\end{multline*}
But $*(l\wedge m)=-i(l\wedge m)$
(see (\ref{polarization of l wedge m}))
and $*(m\wedge\bar m)=+i(l\wedge n)$ so the above formula becomes
\[
L_\mathrm{Weyl}(\xi)
=-\frac18(
l\wedge n\wedge dl
-l\wedge\bar m\wedge dm
-l\wedge m\wedge d\bar m
).
\]
Comparing with (\ref{Lagrangian in complex form}) we arrive at
(\ref{relation between two Lagrangians}). $\square$

\section{Conformal invariance}
\label{Conformal invariance}

Until now we kept the metric fixed but now we shall scale the metric as
(\ref{rescaling metric}) and the Pauli matrices as
\begin{equation}
\label{Conformal invariance equation 1}
\sigma_\alpha\mapsto e^h\sigma_\alpha.
\end{equation}
Recall that here $h:M\to\mathbb{R}$ is an arbitrary scalar function.
Let us also scale the spinor field as
\begin{equation}
\label{Conformal invariance equation 2}
\xi\mapsto e^{-(3/2)h}\xi.
\end{equation}
It is well know that the Weyl Lagrangian
(\ref{Weyl's Lagrangian}) is invariant under the transformation
(\ref{rescaling metric}),
(\ref{Conformal invariance equation 1}),
(\ref{Conformal invariance equation 2}).

Examination of formulae
(\ref{relationship between coframe and spinor field}),
(\ref{second order Pauli matrices})
shows that the transformation
(\ref{rescaling metric}),
(\ref{Conformal invariance equation 1}),
(\ref{Conformal invariance equation 2})
induces the following transformation of the complex coframe
(\ref{definition of covector field l}),
(\ref{definition of covector field m}),
(\ref{definition of covector field n}):
\begin{equation}
\label{Conformal invariance equation 3}
\begin{pmatrix}
l\\ m\\ n
\end{pmatrix}
\mapsto
\begin{pmatrix}
e^{-2h}l\\ e^{h}m\\ e^{4h}n
\end{pmatrix}
\end{equation}
Of course, it is easy to check directly that our teleparallel Lagrangian
(\ref{Lagrangian in complex form})
is invariant under the transformation (\ref{Conformal invariance equation 3}).

The transformation (\ref{Conformal invariance equation 3}) is a
composition of two commuting transformations: a conformal rescaling
of the coframe (\ref{rescaling coframe})
and a Lorentz boost
\[
\begin{pmatrix}
\vartheta^0\\
\vartheta^3
\end{pmatrix}
\mapsto
\begin{pmatrix}
\cosh3h&-\sinh3h\\
-\sinh3h&\cosh3h
\end{pmatrix}
\begin{pmatrix}
\vartheta^0\\
\vartheta^3
\end{pmatrix}.
\]
The presence of a Lorentz boost
in this argument
is somewhat unnatural so we suggest below
a modified version of our teleparallel Lagrangian,
one for which conformal invariance is self-evident.
Recall that our original teleparallel Lagrangian
$L_\mathrm{tele}(\vartheta)$
was defined by formula
(\ref{teleparallel Lagrangian}) or, equivalently,
in terms of the complex coframe, by formula
(\ref{Lagrangian in complex form}).

Put
\begin{equation}
\label{Conformal invariance equation 4}
\tilde L_\mathrm{tele}(\vartheta,s):=
sL_\mathrm{tele}(\vartheta)
=sl\wedge T^\mathrm{ax}
=(s/6)\ l\wedge(n\wedge dl-\bar m\wedge dm-m\wedge d\bar m)
\end{equation}
where $s:M\to(0,+\infty)$ is a scalar function. The function $s$ will
play the role of an additional dynamical variable.
In view of (\ref{rescaling axial torsion})
the Lagrangian (\ref{Conformal invariance equation 4}) does not change
if we scale
the coframe as (\ref{rescaling coframe}),
the metric as (\ref{rescaling metric})
and the scalar $s$ as
$s\mapsto e^{-3h}s$.
Hence, the Lagrangian (\ref{Conformal invariance equation 4}) is conformally
invariant and, moreover, this conformal invariance is quite obvious.

Let us now examine the properties of the Lagrangian
(\ref{Conformal invariance equation 4}) for \emph{fixed} metric.
Of course, it is invariant under the action of the group $H$
which was described implicitly in Section \ref{Main result}
and explicitly in Section \ref{The gauge group H}
(see formula (\ref{The gauge group H equation 4})).
However, it is also invariant under the transformation
\begin{equation}
\label{Conformal invariance equation 5}
\begin{pmatrix}
l\\ m\\ n\\ s
\end{pmatrix}
\mapsto
\begin{pmatrix}
e^{-k}l\\ m\\ e^kn\\ e^ks
\end{pmatrix}
\end{equation}
where $k:M\to\mathbb{R}$ is an arbitrary scalar function.
The transformation (\ref{Conformal invariance equation 4}) is a
composition of two transformations: a Lorentz boost
\[
\begin{pmatrix}
\vartheta^0\\
\vartheta^3
\end{pmatrix}
\mapsto
\begin{pmatrix}
\cosh k&-\sinh k\\
-\sinh k&\cosh k
\end{pmatrix}
\begin{pmatrix}
\vartheta^0\\
\vartheta^3
\end{pmatrix}
\]
and a rescaling of the scalar $s$, $s\mapsto e^ks$. We will denote
the infinite-dimensional
Lie group of transformations (\ref{Conformal invariance equation 5})
by $J(M)$.

Thus, having incorporated into our original teleparallel Lagrangian
(\ref{teleparallel Lagrangian})
an additional dynamical variable, the positive scalar function $s$,
we have acquired an additional gauge degree of freedom. The new
(extended) gauge group is
\begin{multline*}
\tilde H
=H\ltimes J(M)
=(B^2(M)\ltimes\mathrm{U}(1))\ltimes J(M)
\\
=(B^2(M)\ltimes J(M))\ltimes\mathrm{U}(1)
=B^2(M)\ltimes(J(M)\times\mathrm{U}(1))
\end{multline*}
where the symbol ``$\ltimes$'' stands for the semidirect product
with the normal subgroup coming first.
The action of $\tilde H$ preserves the 2-form $l\wedge m$ modulo
$\mathrm{U}(1)$ and modulo rescaling by a positive scalar function.

We have established the following analogue of
Theorem \ref{main theorem 1}.

\begin{theorem}
\label{main theorem 3}
The modified teleparallel Lagrangian
(\ref{Conformal invariance equation 4}) is invariant under the action of the
group $\tilde H$.
\end{theorem}

In view of Theorem \ref{main theorem 3} we call two sets of
dynamical variables ``coframe + positive scalar'' equivalent if they differ
by a transformation from the group $\tilde H$ and gather sets of
dynamical variables into equivalence classes according to this
relation.
The following is an analogue of
Theorem \ref{main theorem 2}.

\begin{theorem}
\label{main theorem 4}
The equivalence classes of coframes $\vartheta$ and positive scalars $s$
on the one hand and nonvanishing spinor fields
$\xi$ on the other are in a one-to-one correspondence given by the formula
\begin{equation}
\label{relationship between coframe and spinor field with scalar}
s\left(l\wedge m\right)_{\alpha\beta}
\stackrel{\operatorname{mod}\mathrm{U}(1)}=
\sigma_{\alpha\beta ab}\xi^a\xi^b
\end{equation}
where $l$ and $m$ are defined by formulae
(\ref{definition of covector field l})
and
(\ref{definition of covector field m})
respectively,
$\vartheta$, $s$ and $\xi$ are arbitrary representatives of
the corresponding equivalence classes and
$\sigma_{\alpha\beta}$ are ``second order'' Pauli matrices
(\ref{second order Pauli matrices}).
Furthermore, under the
correspondence~(\ref{relationship between coframe and spinor field with scalar})
we have
\begin{equation}
\label{relation between two Lagrangians with scalar}
\tilde L_\mathrm{tele}(\vartheta,s)=-\frac43\,L_\mathrm{Weyl}(\xi).
\end{equation}
\end{theorem}

The proof of the first part of Theorem \ref{main theorem 4} (formula
(\ref{relationship between coframe and spinor field with scalar}))
is essentially a repetition of the proof of the first
part of Theorem~\ref{main theorem 2}:
take argument from the beginning of Section \ref{Proof of Theorem 1}
and add one gauge degree of freedom.

As to the second part of Theorem \ref{main theorem 4} (formula
(\ref{relation between two Lagrangians with scalar})), it simply follows
from the second part of Theorem \ref{main theorem 2} (formula
(\ref{relation between two Lagrangians})). Indeed, when we replace
(\ref{relationship between coframe and spinor field})
by (\ref{relationship between coframe and spinor field with scalar})
the spinor field scales as $\xi\mapsto\sqrt s\,\xi$. But
\[
-\frac43\,L_\mathrm{Weyl}(\sqrt s\,\xi)
=-\frac43\,sL_\mathrm{Weyl}(\xi)
\stackrel{\text{by (\ref{relation between two Lagrangians})}}=
sL_\mathrm{tele}(\vartheta)
\stackrel{\text{by (\ref{Conformal invariance equation 4})}}=
L_\mathrm{tele}(\vartheta,s)
\]
giving us (\ref{relation between two Lagrangians with scalar}).

\section{Discussion}
\label{Discussion}

Throughout the paper we dealt with Weyl's Lagrangian (\ref{Weyl's Lagrangian})
as opposed to Weyl's equation (\ref{Weyl's equation})
For Weyl's Lagrangian we found a simple teleparallel
representation~(\ref{teleparallel Lagrangian}).
If one wishes to rewrite Weyl's
equation in teleparallel form then one has to vary the action
with respect to the coframe $\vartheta$
and the resulting teleparallel representation of Weyl's equation
does not turn out to be that simple, the reason being that in
performing the variation one has to maintain the metric constraint
(\ref{constraint for coframe}) because we agreed
(see first paragraph of Section \ref{Main result})
to keep the metric fixed (prescribed).
The corresponding calculations are carried out in
Appendix \ref{Weyl's equation in teleparallel form}.

The teleparallel representation of Weyl's equation was first derived
by Griffiths and Newing \cite{MR0332092}. Our contribution is the
teleparallel representation of Weyl's Lagrangian and observation
that for the Lagrangian things become much simpler.

Now, formula (\ref{relation between two Lagrangians})
(as well as its generalised version
(\ref{relation between two Lagrangians with scalar}))
holds for \emph{any} Lorentzian metric so when using this formula
there is really no need
in assuming the metric to be fixed. Say, one can
vary the action with respect to the metric~$g$
to derive the teleparallel representation of the neutrino energy--momentum tensor.
The calculations are quite straightforward but we do not perform them in this
paper for the sake of brevity.

Let us now examine the geometric meaning of the covector field $l$
defined by formula (\ref{definition of covector field l}).
If we choose Pauli matrices in the special way
(\ref{special formula for Pauli matrices}), (\ref{Pauli matrices s})
we get (\ref{special formula for spinor}) which immediately implies
\begin{equation}
\label{neutrino current}
l_\alpha=\sigma_{\alpha a\dot b}\xi^a\bar\xi^{\dot b}.
\end{equation}
Formula (\ref{neutrino current}) remains true for any choice of
Pauli matrices because its RHS has an invariant meaning. More
specifically, the RHS of (\ref{neutrino current}) is the well-known
expression for the neutrino current. In light of this it is not surprising
that our field equations imply that the divergence of $l$ is zero,
see formula (\ref{complexification 1}) in Appendix
\ref{Weyl's equation in teleparallel form}.

The main issue with our model is that our Lagrangian
(\ref{teleparallel Lagrangian}) (as well as its generalised version
(\ref{Conformal invariance equation 4})) is not invariant under
rigid Lorentz transformations of the coframe. In the remainder of
this section we sketch out a way of dealing with this issue.

Consider the Lagrangian
\begin{equation}
\label{quadratic teleparallel Lagrangian}
L(\vartheta,s):=s\|T^\mathrm{ax}\|^2*1
\end{equation}
where $s:M\to(0,+\infty)$ is a scalar function which plays the role
of an additional dynamical variable. This Lagrangian is Lorentz
invariant and is a special case of a general quadratic Lorentz
invariant Lagrangian (a general one contains squares of all three
irreducible pieces of torsion). The special feature of the
Lagrangian (\ref{quadratic teleparallel Lagrangian}) is that it is
conformally invariant: it does not change if we rescale the coframe as
(\ref{rescaling coframe}) and the scalar $s$ as
$s\mapsto e^{-2h}s$.

Of course, a positive scalar $s$ is equivalent to a positive density
$\rho$:
\linebreak
$\rho=s\sqrt{|\det g|}$. Thus, having the scalar function
$s$ as a dynamical variable is equivalent to having the density
$\rho$ as a dynamical variable. Thinking in terms of an unknown density
$\rho$ is more natural from the physical viewpoint. However, in this paper we will
stick with the scalar $s$.

We vary the action $S(\vartheta,s):=\int L(\vartheta,s)$
with respect to the scalar $s$ and
with respect to the coframe $\vartheta$ subject to the metric constraint
(\ref{constraint for coframe}), which gives us the
Euler--Lagrange field equations.
The fundamental difference between our original conformally invariant Lagrangian
(\ref{Conformal invariance equation 4}) and the new conformally invariant Lagrangian
(\ref{quadratic teleparallel Lagrangian})
is that the latter is quadratic in torsion, hence the field equations
for (\ref{quadratic teleparallel Lagrangian}) will be second order.

Suppose now that the metric is Minkowski.
It turns our that in this case one can construct an explicit
solution of the field equations
for (\ref{quadratic teleparallel Lagrangian}).
This construction goes as follows.

Let $l\ne0$ be a constant real lightlike covector lying on the
forward light cone and let $\bm{\vartheta}$ be a constant coframe
such that
$l\perp\bm{\vartheta}^1$,
$l\perp\bm{\vartheta}^2$;
here ``constant'' means ``parallel with respect to the
Levi-Civita connection induced by the Minkowski metric''.
Then, of course,
\begin{equation}
\label{definition of covector field l bold}
l=c(\bm{\vartheta}^0+\bm{\vartheta}^3)
\end{equation}
where $c>0$ is some constant (compare with formula
(\ref{definition of covector field l})).
Put
\begin{equation}
\label{plane wave}
\begin{pmatrix}
\vartheta^0\\
\vartheta^1\\
\vartheta^2\\
\vartheta^3
\end{pmatrix}:=
\begin{pmatrix}
1&0&0&0\\
0&\cos2\varphi&\pm\sin2\varphi&0\\
0&\mp\sin2\varphi&\cos2\varphi&0\\
0&0&0&1
\end{pmatrix}
\begin{pmatrix}
\bm{\vartheta}^0\\
\bm{\vartheta}^1\\
\bm{\vartheta}^2\\
\bm{\vartheta}^3
\end{pmatrix},
\qquad s=\operatorname{const}>0
\end{equation}
where $\varphi:=\int l\cdot dx$ and $x^\alpha$ are local
coordinates. Straightforward calculations show that this coframe
$\vartheta$ and scalar $s$ are indeed a solution of the field
equations for (\ref{quadratic teleparallel Lagrangian}). We call
this solution a \emph{plane wave} with \emph{momentum} $l$. The
upper sign in (\ref{plane wave}) corresponds to the massless
neutrino and lower sign corresponds to the massless antineutrino.
Note that we can distinguish the neutrino from the antineutrino
without resorting to negative energies. Note also that we
automatically get only one type of neutrino (left-handed) and one
type of antineutrino (right-handed).

Suppose now that we are seeking solutions which are not necessarily
plane waves. This can be done using perturbation theory. In the
language of spinors perturbation means that we assume the spinor
field to be of the form
``slowly varying spinor $\times\ e^{-i\varphi}$''.
We claim that application of a perturbation argument reduces the
quadratic (in torsion) Lagrangian (\ref{quadratic teleparallel Lagrangian})
to the linear (in torsion) Lagrangian (\ref{Conformal invariance equation 4}).
At the most basic level this can be explained as follows.
Note that for a plane wave we have the following two identities:
$T^\mathrm{ax}=\pm\frac43*l$ and $l=c(\vartheta^0+\vartheta^3)$
(compare the latter with (\ref{definition of covector field l bold})).
Thus, for a plane wave we have
\begin{equation}
\label{axial torsion of plane wave}
T^\mathrm{ax}=\pm\frac43c*(\vartheta^0+\vartheta^3).
\end{equation}
We now linearize (in torsion)
the quadratic Lagrangian (\ref{quadratic teleparallel Lagrangian})
about the point~(\ref{axial torsion of plane wave}).
We get, up to a constant factor,
the linear Lagrangian~(\ref{Conformal invariance equation 4}).

The bottom line is that we believe that the true Lagrangian
of a massless neutrino field is the quadratic Lagrangian
(\ref{quadratic teleparallel Lagrangian}).
The linear Lagrangian (\ref{Conformal invariance equation 4})
(which is equivalent to Weyl's Lagrangian (\ref{Weyl's Lagrangian}))
arises only if one adopts the perturbative approach.

The detailed analysis of the quadratic Lagrangian
(\ref{quadratic teleparallel Lagrangian})
will be the subject of a separate paper.
Elements of this analysis have been performed in
\cite{mygrossman2006,rome}.

\section*{Acknowledgments}

The authors are grateful to C.~B\"ohmer and F.~W.~Hehl for helpful advice.

\appendix

\section{Brief introduction to teleparallelism}
\label{A brief introduction to teleparallelism}

Given a coframe $\vartheta$, we introduce a covariant derivative
$|\nabla|$ such that
$|\nabla|\vartheta=0$.
We repeat this formula
giving frame and tensor indices explicitly:
$|\nabla|_\alpha\vartheta^j_\beta=0$.
We then rewrite the formula
in even more explicit form:
\begin{equation}
\label{A brief introduction to teleparallelism equation 1}
\partial_\alpha\vartheta^j_\beta
-|\Gamma|^\gamma{}_{\alpha\beta}\vartheta^j_\gamma=0
\end{equation}
where $|\Gamma|^\gamma{}_{\alpha\beta}$ are the connection coefficients.
Note that formula
(\ref{A brief introduction to teleparallelism equation 1})
has three free indices $j$, $\alpha$, $\beta$
running through the values $0,1,2,3$.
Note also that the connection coefficient
$|\Gamma|^\gamma{}_{\alpha\beta}$ has three indices $\alpha$, $\beta$, $\gamma$
running through the values $0,1,2,3$.
Hence,
(\ref{A brief introduction to teleparallelism equation 1})
can be viewed as a system of 64 inhomogeneous linear algebraic equations
for the determination of the 64 unknown connection coefficients
$|\Gamma|^\gamma{}_{\alpha\beta}$. It is easy to see that its unique solution is
\begin{equation}
\label{A brief introduction to teleparallelism equation 2}
|\Gamma|^\gamma{}_{\alpha\beta}=
o_{ik}g^{\gamma\delta}
\vartheta^i_\delta\partial_\alpha\vartheta^k_\beta.
\end{equation}
The corresponding connection is called
\emph{teleparallel}.
When writing the teleparallel covariant derivative
and connection coefficients we use the ``modulus'' sign to distinguish
these from the Levi-Civita covariant derivative
and connection coefficients for which we use curly brackets.

Thus, we have two different connections: the Levi-Civita connection
used in the main text of the paper and the teleparallel connection
used in this appendix.
Both are metric compatible: $\{\nabla\}g=|\nabla|g=0$.
The Levi-Civita connection is uniquely
determined by the metric whereas the teleparallel connection is
uniquely determined by the coframe. For the Levi-Civita connection
torsion is zero whereas for the teleparallel connection curvature is
zero. Thus, in a sense, the Levi-Civita and teleparallel connections
are antipodes.

``Teleparallelism'' stands for ``distant parallelism''. What is
meant here is that the result of parallel transport of a vector (or
a covector) does not depend on the choice of curve connecting the
two points. This fact can be expressed in even simpler terms as
follows. Suppose we have two covectors, $u$ and $v$, at two
different points, $P$ and $Q$, of our manifold (spacetime)
$M$. We need to establish whether $u$ and $v$ are parallel.
To do this, we use the coframe as a basis and write
$u=a_j\vartheta^j$, $v=b_j\vartheta^j$. By definition, the covectors
$u$ and $v$ are said to be parallel if $a_j=b_j$.

Formula (\ref{A brief introduction to teleparallelism equation 2})
allows us to evaluate torsion of the teleparallel connection:
\[
T^\gamma{}_{\alpha\beta}
:=|\Gamma|^\gamma{}_{\alpha\beta}-|\Gamma|^\gamma{}_{\beta\alpha}
=o_{ik}g^{\gamma\delta}
\vartheta^i_\delta
(\partial_\alpha\vartheta^k_\beta-\partial_\beta\vartheta^k_\alpha)
=o_{ik}g^{\gamma\delta}
\vartheta^i_\delta
(d\vartheta^k)_{\alpha\beta}
\]
where $\,d\,$ denotes the exterior derivative.
Lowering the first tensor index gives a neater representation
$T_{\gamma\alpha\beta}=o_{ik}
\vartheta^i_\gamma
(d\vartheta^k)_{\alpha\beta}$.
Dropping tensor indices altogether we get
\begin{equation}
\label{A brief introduction to teleparallelism equation 3}
T=o_{ik}
\vartheta^i\otimes d\vartheta^k.
\end{equation}

It is know \cite{MR1340371,MR1925542,cartantorsionreview}
that torsion decomposes into three irreducible pieces called tensor torsion,
vector torsion and axial torsion. (Vector torsion is sometimes called trace
torsion.) In this paper we use only the axial piece. Axial torsion has a very
simple meaning: it is the totally antisymmetric piece
$T^\mathrm{ax}{}_{\alpha\beta\gamma}
=\frac13
(T_{\alpha\beta\gamma}
+T_{\gamma\alpha\beta}
+T_{\beta\gamma\alpha})$.
Substituting (\ref{A brief introduction to teleparallelism equation 3})
into this general formula we arrive at (\ref{axial torsion}).

Of course, there is much more to teleparallelism than the elementary
facts sketched out above.
Modern reviews of the physics of teleparallelism can be found in
\cite{MR583723,MR1661422,MR1617858,deandrade-2000,MR1871425,MR1976711}.

\section{Weyl's equation in teleparallel form}
\label{Weyl's equation in teleparallel form}

In this appendix we write down explicitly the Euler--Lagrange field equations
resulting from the variation of the action
\begin{equation}
\label{teleparallel action}
S_\mathrm{tele}:=\int L_\mathrm{tele}
=\int l\wedge T^\mathrm{ax}
=\frac13\,p_io_{jk}\int\vartheta^i\wedge\vartheta^j\wedge d\vartheta^k
\end{equation}
with respect to the coframe $\vartheta$ subject to the metric constraint
(\ref{constraint for coframe}). Here by $p_i$ we denote the quartet of constants
$p_i:=\begin{pmatrix}1&0&0&1\end{pmatrix}$.

The variation of the coframe is given by the formula
\begin{equation}
\label{variation of the coframe}
\delta\vartheta^j{}_k=F^j{}_k\vartheta^k
\end{equation}
where the $F^j{}_k$ are real scalar functions satisfying the antisymmetry condition
\begin{equation}
\label{antisymmetry of F}
F_{jk}=-F_{kj}.
\end{equation}
Condition (\ref{antisymmetry of F}) ensures that the variation of the RHS
of (\ref{constraint for coframe}) is zero. Of course, the $\Lambda^j{}_k$
appearing the RHS of (\ref{Lorentz transformation 1}) are expressed via the
$F^j{}_k$ as
\[
\Lambda^j{}_k=\delta^j_k+F^j{}_k+\frac12F^j{}_lF^l{}_k+\ldots
\]
(exponential series), or, in matrix notation, $\Lambda=e^F$.
Hence, the matrix-function $F$ is the linearization of
the Lorentz transformation $\Lambda$ about the identity.

Substituting (\ref{variation of the coframe}) into
(\ref{teleparallel action}) we get
\[
3\delta S_\mathrm{tele}
=p_io_{jk}\int(
F^i{}_l\vartheta^l\wedge\vartheta^j\wedge d\vartheta^k
+F^j{}_l\vartheta^i\wedge\vartheta^l\wedge d\vartheta^k
+F^k{}_l\vartheta^i\wedge\vartheta^j\wedge d\vartheta^l
+\vartheta^i\wedge\vartheta^j\wedge dF^k{}_l\wedge\vartheta^l)
\]
where $dF^k{}_l$ is the gradient of the scalar function $F^k{}_l$.
Upon contraction with $o_{jk}$ the second and third terms in the
integrand cancel out in view of (\ref{antisymmetry of F}) (that this
would happen was clear a priori because axial torsion is invariant
under rigid Lorentz transformations) so the above formula becomes
\[
3\delta S_\mathrm{tele}
=\int(
p^io_{lk}F_{ij}\vartheta^j\wedge\vartheta^l\wedge d\vartheta^k
+p_ko_l{}^i\vartheta^k\wedge\vartheta^l\wedge dF_{ij}\wedge\vartheta^j)
\]
where $p^i:=o^{ij}p_j$.
Integration by parts and antisymmetrization in $i$, $j$ gives
\[
6\delta S_\mathrm{tele}
=\int F_{ij}
(p^io_{lk}\vartheta^j\wedge\vartheta^l\wedge d\vartheta^k
-p^jo_{lk}\vartheta^i\wedge\vartheta^l\wedge d\vartheta^k
-2p_kd(\vartheta^k\wedge\vartheta^i\wedge\vartheta^j)).
\]
Thus, our field equations are
\begin{equation}
\label{teleparallel field equations}
p^io_{lk}\vartheta^j\wedge\vartheta^l\wedge d\vartheta^k
-p^jo_{lk}\vartheta^i\wedge\vartheta^l\wedge d\vartheta^k
-2p_kd(\vartheta^k\wedge\vartheta^i\wedge\vartheta^j)=0.
\end{equation}

The field equations (\ref{teleparallel field equations})
are, of course, equivalent to
\begin{equation}
\label{teleparallel field equations with Hodge star}
*[p^io_{lk}\vartheta^j\wedge\vartheta^l\wedge d\vartheta^k
-p^jo_{lk}\vartheta^i\wedge\vartheta^l\wedge d\vartheta^k
-2p_kd(\vartheta^k\wedge\vartheta^i\wedge\vartheta^j)]=0.
\end{equation}
The advantage of the representation
(\ref{teleparallel field equations with Hodge star}) is that the left-hand
sides of (\ref{teleparallel field equations with Hodge star}) are scalars
and not 4-forms as in (\ref{teleparallel field equations}).
We denote the left-hand
sides of (\ref{teleparallel field equations with Hodge star}) by
$G^{ij}$. Note the antisymmetry $G^{ij}=-G^{ji}$.

We will now rewrite our field equations
(\ref{teleparallel field equations with Hodge star})
in more compact form
in terms of the complex coframe
(\ref{definition of covector field l}),
(\ref{definition of covector field m}),
(\ref{definition of covector field n}).

We note first that
$G^{12}=4\{\nabla\}_\alpha l^\alpha$.
Thus, our field equations
(\ref{teleparallel field equations with Hodge star})
imply
\begin{equation}
\label{complexification 1}
\{\nabla\}_\alpha l^\alpha=0.
\end{equation}
Note that the scalar $G^{03}$ also has a clear geometric meaning:
$G^{03}=3*L_\mathrm{tele}$.

Put
\[
q_j:=\begin{pmatrix}0&1&i&0\end{pmatrix},
\qquad
r_j:=\begin{pmatrix}1&0&0&-1\end{pmatrix},
\]
\[
A_{jk}:=p_jq_k-p_kq_j,\qquad
B_{jk}:=p_jr_k-p_kr_j-q_j\bar q_k+q_k\bar q_j,\qquad
C_{jk}:=r_j\bar q_k-r_k\bar q_j.
\]
The antisymmetric matrices
$\operatorname{Re}A$, $\operatorname{Im}A$,
$\operatorname{Re}B$, $\operatorname{Im}B$,
$\operatorname{Re}C$, $\operatorname{Im}C$
are linearly independent, therefore the system of 6 real equations
(\ref{teleparallel field equations with Hodge star}) is equivalent to the
system of 3 complex equations
\[
A_{ij}G^{ij}=0,\qquad
B_{ij}G^{ij}=0,\qquad
C_{ij}G^{ij}=0.
\]
Straightforward calculations show that $A_{ij}G^{ij}$ is
zero for any coframe~$\vartheta$
(this is actually a consequence of Theorem \ref{main theorem 1}),
hence our real field equations
(\ref{teleparallel field equations with Hodge star})
are equivalent to the pair of complex equations
\begin{equation}
\label{complexification 2}
B_{ij}G^{ij}=0,
\qquad
C_{ij}G^{ij}=0.
\end{equation}

As the systems
(\ref{teleparallel field equations with Hodge star})
and
(\ref{complexification 2})
are equivalent
and as equation (\ref{complexification 1}) is a consequence of
(\ref{teleparallel field equations with Hodge star}),
equation (\ref{complexification 1}) is also a consequence of
(\ref{complexification 2}).
Hence we can extend the system
(\ref{complexification 2}) by adding equation (\ref{complexification 1}):
the system (\ref{complexification 2}) is equivalent to the system
(\ref{complexification 2}), (\ref{complexification 1}).
The advantage of having (\ref{complexification 1}) as a separate equation
is that it simplifies subsequent calculations.

We now examine our system of field equations
(\ref{complexification 2}), (\ref{complexification 1}).
Straightforward calculations with account of (\ref{complexification 1})
give
\[
B_{ij}G^{ij}=-8i\bar m^\alpha v_\alpha,
\qquad
C_{ij}G^{ij}=8in^\alpha\bar v_\alpha
\]
where
\begin{equation}
\label{explicit formula for v}
v_\alpha:=
\{\nabla\}^\beta(l\wedge m)_{\alpha\beta}-m^\beta\{\nabla\}_\alpha l_\beta.
\end{equation}
Thus, our system of field equations
(\ref{complexification 2}), (\ref{complexification 1})
is equivalent to
\begin{equation}
\label{complexification 3}
\bar m^\alpha v_\alpha=0,
\qquad
n^\alpha v_\alpha=0
\end{equation}
and (\ref{complexification 1}).
But $\operatorname{Re}(\bar m^\alpha v_\alpha)=2\{\nabla\}_\alpha l^\alpha$,
so (\ref{complexification 1}) is a consequence of (\ref{complexification 3}).
Hence, (\ref{complexification 3}) is the full system
of field equations. It is equivalent to the original
system of field equations (\ref{teleparallel field equations with Hodge star}).

It is easy to see that for any coframe $\vartheta$ we have
\begin{equation}
\label{complexification 4}
m^\alpha v_\alpha=0,
\qquad
l^\alpha v_\alpha=0
\end{equation}
so the pair of scalar complex equations (\ref{complexification 3}) is equivalent
to the complex covector equation
\begin{equation}
\label{complexification 5}
v=0.
\end{equation}
Recall that the LHS of this equation is defined by formula
(\ref{explicit formula for v}).

Equation (\ref{complexification 5})
is the compact ``tetrad'' representation of the Weyl equation
found by Griffiths and Newing \cite{MR0332092}.
Griffiths and Newing derived (\ref{complexification 5}) directly from
Weyl's equation (\ref{Weyl's equation}), without examining the Weyl Lagrangian
(\ref{Weyl's Lagrangian}).

Let us have a closer look at equation  (\ref{complexification 5}) so as to establish
the actual number of independent ``scalar'' equations contained in it
and the actual number of independent ``scalar'' unknowns.
It would seem that (\ref{complexification 5}) is a system of 4
complex ``scalar'' equations (4 being the number of components of
the covector~$v$) for 6 real ``scalar'' unknowns (6 being the dimension of the
Lorentz group).
But we already know that we a priori have identities
(\ref{complexification 4})
so equation (\ref{complexification 5}) is equivalent to the pair of
scalar complex equations (\ref{complexification 3}).
It is also easy to see that $v$ is invariant under the action of
the transformation (\ref{The gauge group H equation 3}),
hence the set of solutions to equation (\ref{complexification 5})
is invariant under this transformation
which means that we are dealing with a pair of complex ``scalar''
unknowns (see argument in the beginning of Section
\ref{Proof of Theorem 2}).
Thus, equation (\ref{complexification 5}) is a system of 2
complex ``scalar'' equations for 2 complex ``scalar'' unknowns,
as expected of the Weyl equation.

Note that the scalar $\bar m^\alpha v_\alpha$
appearing in the LHS of (\ref{complexification 3})
is also invariant
under the action of the transformation (\ref{The gauge group H equation 3})
and can be written down explicitly as
$\bar m^\alpha v_\alpha
=2\{\nabla\}_\alpha l^\alpha-\frac{3i}2*L_\mathrm{tele}$.

\end{document}